\documentclass[12pt,a4paper]{article}

\usepackage{verbatim}
\usepackage{fancyvrb}
\usepackage{graphicx}
\usepackage[numbers]{natbib}
\usepackage{latexsym,amssymb,amsmath,amsfonts}
\usepackage{lmodern}	
\usepackage[usenames]{xcolor}
\usepackage{multirow}
\usepackage{siunitx}
\usepackage{float}
\usepackage[autostyle]{csquotes}
\usepackage[paper=a4paper,top=1in,bottom=1in,margin=1in]{geometry}
\usepackage{ragged2e}

\usepackage{subcaption}

\usepackage{fancyhdr}
\usepackage{hyperref}
\usepackage{pdflscape}
\usepackage{setspace}

\usepackage{enumitem}
\setlist[itemize]{noitemsep,topsep=0pt}

\usepackage{lastpage}

\definecolor{javared}{rgb}{0.6,0,0} 
\definecolor{javagreen}{rgb}{0.25,0.5,0.35} 
\definecolor{javapurple}{rgb}{0.5,0,0.35} 
\definecolor{javadocblue}{rgb}{0.25,0.35,0.75} 
\usepackage{listings}
\begin{document}

\setlength{\topmargin}{-0.5cm} 
\addtolength{\textheight}{1.5cm} 
\begin{titlepage}
\newgeometry{top=1in,bottom=1in,right=1in,left=1in}



\begin{flushright}
\includegraphics[width=0.3\textwidth]{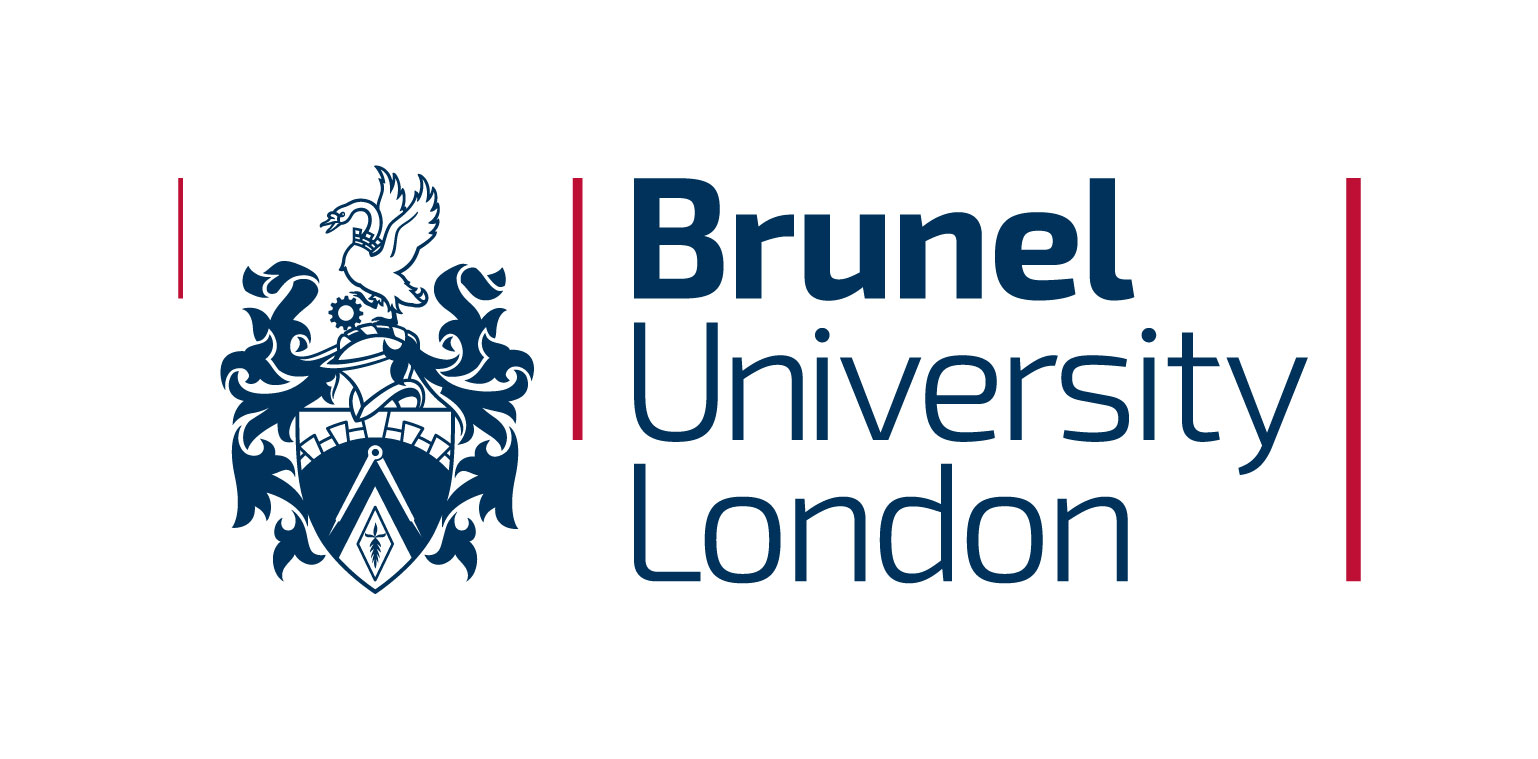}
\end{flushright}
\center 
\Large{\textbf{College of Engineering, Design and Physical Sciences}}

\vspace{0.2cm}

\large{\textbf{Department of Electronic \& Computer Engineering}}

\vspace{0.2cm}

\large{MSc Distributed Computing Systems Engineering}

\vspace{1.4cm}

\LARGE{\textbf{Brunel University London}}

\vspace{2cm}

{ \huge \bfseries Adaptive Event Dispatching in Serverless Computing
Infrastructures}

\vspace{2cm}

\textsc{\LARGE Interim Report}

\vspace{3.6cm}

\begin{minipage}{\textwidth}
\center
\large \bfseries 
\begin{tabular}{l l}
Student: & Manuel Stein (1644641)\\
Supervisor: & Prof. Maozhen Li\\
Date: & September 29, 2017 \\ 
\end{tabular}
\end{minipage}

\vfill 

\end{titlepage}

\newgeometry{top=1in,bottom=1in,right=20mm,left=40mm}
\pagestyle{fancy}
\fancyhf{}
\fancyhead[R]{\leftmark}
\fancyfoot[C]{page \thepage}

\tableofcontents

\doublespacing
\section{Introduction}
Serverless computing is an emerging service model in distributed computing
systems. The term captures cloud-based event-driven distributed application
design and stems from its completely resource-transparent deployment model, i.e.
serverless. This work thesisizes that adaptive event dispatching can improve
current serverless platform resource efficiency by considering locality and
dependencies. These design contemplations have also been formulated by
Hendrickson et al in \cite{openlambda}, which identifies the requirement that
\enquote{Serverless load balancers must make low-latency decisions while
considering session, code and data locality}. This interim report investigates
the economical importance of the emerging trend and asserts that existing
serverless platforms still do not optimize for data locality, whereas a variety
of scheduling methods are available from distributed computing research which
have proven to increase resource efficiency.

This report is structured as follows. Section \ref{sec:background} gives a
thorough background of the topic. The initial survey provided in section
\ref{sec:survey} assesses the economical aspects (\ref{sec:economic}) of
improving serverless event dispatching and asserts the requirement for adaptive
event dispatching in existing serverless platforms (\ref{sec:architectures}).
Objectives for the project are provided in section \ref{sec:objectives}. Section
\ref{sec:methods} discusses methods used in other areas that can be applied
in the design of a novel adaptive event dispatching. The remainder provides a
project task plan in section \ref{sec:timeline} and concludes with the expected
deliverables of the project in section \ref{sec:deliverables}.

\section{Project Background}\label{sec:background}

Serverless has recently emerged as a trend in Cloud computing. Its
Function-as-a-Service model envisions applications to be partitioned
(over-decomposed) and designed as event-driven applications. The event-driven
design paradigm\cite{eventdrivenreactivity}\cite{cloudeventprogramming} has
already been adopted to develop highly elastic components for self-managing
applications\cite{cloudnativeeventbased}. Serverless may give way to reactive
programming for software\cite{reactiveprogrammingsurvey} that would
automatically scale on Cloud platforms. A background on the serverless trend is
given below (\ref{sec:serverless}).

But the trending decomposition of application components into smaller functions
bears the risk of significant overhead because in its plain form, every
invocation is encapsulated with request authorization, event dispatching, code
loading and heavy remote data access which raises performance challenges
discussed in section \ref{sec:challenges}. Section \ref{sec:eventdispatching}
explains the pending change from allocation placement to event (task)
scheduling on shared Cloud infrastructures.

\subsection{Serverless Computing}\label{sec:serverless}

Serverless computing has only recently appeared. Within a short series of
announcements, Cloud providers alike have released offerings that facilitate
event-driven programming. Amazon Lambda\cite{amazon} first became generally
available (9/4/15), Google released Cloud Functions\cite{google} (11/02/16), IBM
had announced OpenWhisk\cite{ibm} (22/02/16), and Microsoft announced Azure
Functions\cite{amazon} (31/03/16).
At the same time, research papers emerged at prominent venues, such as the
OpenLambda\cite{openlambda} implementation at HotCloud'16 and a review of Cloud
event programming\cite{cloudeventprogramming} at CLOUD'16. The topic has
recently drawn attention of computer science research worldwide. By 2017, the
First International Workshop on Serverless Computing (WoSC) has been launched,
several national developer conferences are themed around serverless computing
(e.g. Serverless London, O'Reilly Software Architecture Conference), and others
have adopted the topic to their calls for paper (e.g. WS-REST). Apart from the
acclaimed praise, serverless can be considered a serious trend in Cloud
computing.

The serverless hype yields expectations for fully automated
operations\cite{serverlessnoops}, better resource utilization, and is promoted
a programming model for event-driven IoT applications
\cite{devicelessedge}\cite{amazoniot}\cite{iotlearn}.
Following the advancements of continuous integration/contiuous development, the
event-based function programming paradigm fits the software
engineering trend to continuously develop, test and release small software
extensions, because it provides every function with its individual management
lifecycle, thereby making the software development even more transparent to
operations and may help to close remaining gaps in DevOps automation to achieve
the controversial target of NoOps\cite{noops}.

Serverless continues the microservices trend that segments traditional servers
into scalable, distributed, fault-tolerant components. Serverless functions are
considered even smaller than microservices. Their small size makes the boot time
of individual functions very short because it allows for individual hosts to
only load the required code and it allows for fast reconfigurations of the
deployment.

Related trends exist in Big Data parallel and distributed computing systems that
allow to deploy custom user-defined functions. Some distributed systems already
leverage data locality to reduce both communication cost and delay. For example,
Apache Hadoop schedules map tasks at data replica locations of its distributed
file system. Spark schedules and distributes parallelizable stages of a task
(directed acyclic graph) on servers that host the partitions of a resilient
distributed dataset (RDD). In some cases the scheduler may even delay single
tasks in hope for a CPU to free up rather than moving data across the network.
More recently, a Big Data company has announced serverless computing based on
Apache Spark\cite{databricks}.

\subsection{Decomposition Performance Challenges}\label{sec:challenges}

Serverless computing would decompose an application into event-triggered
function code and externalize state, while abstracting completely from the
machine configuration.

Decomposing the code of a program requires the computing infrastructure to
dispatch events across servers. The executing node would then have to
communicate to obtain related code and data, load code, contact other
services during execution and store or return results. This is a severe overhead
as compared to a local function call and puts serverless platforms to a
performance challenge. Moreover, events may be part of distributed workflows.
The application decomposition bears the risk of having any two related events
run on different machines such that they need to synchronize remotely on the
shared context data.

To optimize the performance of event-driven serverless function execution, code
and data would ideally be readily cached at the executing host. In existing
microservice frameworks, a messaging middleware dispatches named events to
execution hosts while a data backend provides means to access shared data.
Here, the application needs to correlate both the request stream and data
accesses to optimize efficiency.

Because serverless functions are required to be stateless, the application state
needs to be externalized. Current serverless frameworks use distributed
key-value stores. These provide high availability and fault-tolerance by
employing replication and key hashing for pseudo-random distribution. However,
this practice counteracts data colocation and is difficult to circumvent without
jeopardizing data load balancing. Also, distributed message queues are employed
for scalable event dispatching. The most prominent is Kafka (for its reliability
and scalability) which effectively has the message producer select a consumer
partition, whereas in other systems, message brokers allow for message routing
based on message contents. Ideally, events should be dispatched to hosts where
required data context is already cached to speed-up function execution, but it
seems that current platforms are not standing up to the performance challenge.

\subsection{Adaptive Event Dispatching}\label{sec:eventdispatching}

In commercial serverless computing environments, various sorts of events can be
used to trigger a function, e.g. the creation, update or deletion of a data
element (data access), a system event, an application event or by explicit
invocation (web requests, messages). The traditional on-demand model (VM and
microservices alike) allocates isolated compute capacity for network-based
services, i.e. service instances are stateful, stationary control flows that
communicate with each other. This notion of long-running instances has had its
implications on placement of allocation requests. For example, \cite{danny}
approaches the Cloud service placement as a capacitated host and network
allocation problem and formalizes it as a combination of the general assignment
and the facility location problem. A different approach \cite{jaime} assumes a
hierarchically organized infrastructure and formalizes a minimum cost mixed-cast
flow problem that can be solved using linear integer programming, because of the
insights into the application's data flow model. A huge body of research covers
online and offline optimization of the VM/service placement
(\cite{placementsurvey}).

Placement decisions are still a part of serverless platforms, only computing
resources are not actively consumed until events are dispatched. Serverless
functions are fully data and execution location-transparent, so the workload
distribution is effectively determined by event dispatching. This makes event
dispatching a key component in workload scheduling that needs to consider the
cost in establishing an execution context when adapting the workload
distribution.
\section{Initial Survey}\label{sec:survey}

The terminology of the emerging \textbf{Serverless} hype still lacks clear
definition as assessed in \cite{wosc} to whether it is a subset of event-driven
computing or a location transparency paradigm. While many implementation efforts
target the commercialization of \textbf{Function-as-a-Service} under the term
serverless, few research literature is available on its platform designs.
Amazon's spearheading offer is called Lambda, which stems from the Lambda
calculus and has led to the term \textbf{Lambda architecture} (used e.g.
by \cite{openlambda}). The paradigm to design Cloud applications to run on
serverless platforms has recently been coined \textbf{Cloud event
programming}\cite{cloudeventprogramming}, while distributed applications that
are designed to self-manage their resource allocation have been branded by the
microservice (container-based) movement as \textbf{cloud-native} applications
(\cite{cloudnative},\cite{cncf}) which has led to the attribution of
\textbf{cloud-native event-based} application
design\cite{cloudnativeeventbased}. Cloud-native is specifically connected to
container-based architectures\cite{cncf}, and containers are in fact the most
widely used isolation mechanism for FaaS, supposedly including all public
serverless offers\cite{amazon}\cite{google}\cite{ibm}\cite{azure} who provide
it as an extension to their infrastructure. Economical aspects for the adoption
of event-driven computing as a Cloud service model are discussed in more detail
below.

Some research argues for performance reasons that containers \enquote{cannot
be the unit of deployment}\cite{enddominance} and that \enquote{container
throughput [\ldots] would not be enough to be
cost-effective}\cite{nextgencloud}. Given the diversity of emerging computing
architectures that share common requirements, \cite{nextgencloud} argues that
research for unified environments are required and recommends the cloud-native
application design\cite{cloudnative}. A unified platform would exceed the scope
of this thesis. Instead, prominent serverless platform architectures are
reviewed below to assert the opportunity to develop and integrate an
improved event dispatching.

\subsection{Economical aspects}\label{sec:economic}

Serverless computing is advertised an evolution of Cloud computing in the
growing market of virtual infrastructure provisioning. The following describes
how serverless fits with Cloud economy.

\subsubsection{Cloud service model}
Serverless computation stems from the Cloud business model to offer resources on
demand. Infrastructure-as-a-Service offerings for time sharing of a physical
server infrastructure has created a large market to rent resources. Serverless
is about changing the business model from reserved capacity to actual
utilization (cmp. \cite{economicimpact}) and provides means to deploy
consumer-created (or acquired) applications, so it categorizes as
Platform-as-a-Service model by the NIST Cloud service model
definition\cite{nist}. Meanwhile, distributed application architecture
components have been commoditized for Cloud deployment, such as databases,
messaging middleware, storage subsystems, networks, firewalls, and load
balancers. Serverless aims to complement this ecosystem by providing lean
deployment of highly customized application logic. But with serverless,
dispatching events or routing service requests is an integral part of the
platform and no longer part of the application, so this thesis categorizes as
serverless platform research but not Cloud application design.


\subsubsection{Cloud economy}
In the Cloud service business, platform commodity components (DB, messaging,
CDN) are available through self-service interfaces and billed in terms of
hardware resource consumption rather than software metrics. Few services have
appeared as exception that were billed by the number of API calls (e.g. former
versions of the Google App Engine). Today, almost all pricing schemes have been
simplified to billing by infrastructure metrics (resource amounts over time).
Likewise, FaaS offerings charge by resource metrics, none of which allows sizing
of the CPUs but pin execution to a single processor\cite{singlethreaded}.
Only memory can be scaled, so compute time is charged by \textbf{memory capacity
over time}, e.g. MB-sec or GB-h. Except with IBM\cite{ibmpricing}, a charge is
incurred for the number of
invocations\cite{amazonpricing}\cite{azurepricing}\cite{googlepricing}.
Additional costs are incurred by complementary use of persistent storage. Adam
Eivy (Solution Architect at Walt Disney) has published a warning
cry\cite{Economics} that resource consumption can backfire for large demand
baselines and create higher cost for the FaaS customer than traditional IaaS
virtual resource rental. Remote data access during execution of a function can
cause active waiting time for which the customer is billed.
This work thesisizes that intelligent dispatching can improve cache hits on
colocated data caches/stores to reduce function execution time and resource
consumption. Improving resource efficiency would reduce resource usage for the
customer and increase throughput of function invocations for the provider. Also,
operational models where Cloud customers deploy self-managed FaaS platforms on
public or private infrastructures would benefit from improved efficiency,
reduced cost and ultimately lower energy consumption.

\subsubsection{Data center utilization}
Serverless is supposed to saturate resources better than allocation-based
virtual server leases. The actual server utilization in data centers is
low\cite{comatose} despite IaaS provider claims that the introduction of Cloud
infrastructures has reduced overall carbon emissions\cite{carbonamazon}.
The reason is simple. The customer is required to allocate a resource capacity
upfront. The provider has to provide the contracted amount of allocated
resources and must not fall short of resources during execution (e.g.
due to overbooking). This business aspect has ruined some of the benefits of
resource sharing because (a) execution is limited to only utilize the allocated
part of a machine even if it had residual capacity and (b) allocated but unused
resources are not available to colocated workloads. IaaS providers have adapted
virtual machine flavors in terms of CPU and memory capacity for harmonic bin
packing \cite{harmonic} for less fragmentation or more optimal resource booking,
but spurred research on demand prediction, online scaling and finding optimal
QoS trade-offs, etc. Up to date, this problem drives enormous efforts to
optimize Cloud resource allocation.

FaaS offers much smaller memory flavors in terms of MB not GB (e.g. 64MB, 128MB,
256MB). The accounted time granularity is also much finer (100ms) with allowed
execution times up to 5 minutes rather than hours of allocation, which
alleviates the problem of blocking unused resources. As such, Serverless
computing may finally increase the utilization of servers in Cloud data centers.

On the other hand, the current way of isolating serverless function execution
may yield much, much higher overhead. The memory flavors become easily utilized
by even a tiny function, because the execution requires JIT compilation (cmp.
\cite{openlambda}), core libraries and may use additional packages. When
executions need to load code and data from storage back-ends to a fresh (cold)
container instance before actually starting computation, the bootstrap overhead
may become significant. The current operation model is to isolate every function
execution in its own container regardless of sequential actions on the same
execution context. Event dispatching to servers based on where
code and context has already been cached may increase resource efficiency but
risks a lower overall resource utilization. An adaptive solution is required to
optimize for efficiency within cost-effective boundaries of resource
utilization.

\subsection{Review of Serverless Architectures}\label{sec:architectures}

Serverless platforms, in essence, need to dispatch the event, execute the
function and store related data (event data, code, policies, state).
Most serverless platform designs so far follow the Cloud application
model and are compartmentalized architectures similar to the 3-tier web
architecture that separates the web front-end, an application server and a data
back-end. Merely, the novel Cloud application designs extend this concept with
resilience and elasticity of services \cite{cloudnative} and use resource pooling and
caching. Besides the public offers (AWS Lambda, Google Cloud Function, Microsoft
Azure Functions and IBM Bluemix OpenWhisk), over 20 Function-as-a-Service
platforms have emerged, mostly based on Docker container isolation. Hendrickson
et al.\cite{openlambda} have identified the need for research on load balancing
that considers code and data colocation. The following reviews prominent existing
system designs to assert that adaptive event dispatching is still missing from
todays platforms.

\subsubsection{OpenLambda}

Hendrickson et al. have published the OpenLambda platform \cite{openlambda} to
\enquote{facilitate research on Lambda architectures}. The group has identified
colocation awareness as a requirement for load balancing. The project initially
used the front-end proxy to load balance events across worker
instances\cite{openlambdaslides} which keeps forwarding latency short. The group
published ideas early but had only round robin load balancing and worked on a
balancer using gRPC \cite{openlambdalb}. Unfortunately, the group has ceased
development.

\subsubsection{IBM OpenWhisk}

\begin{figure}[H]
\centering
\includegraphics[width=0.8\textwidth]{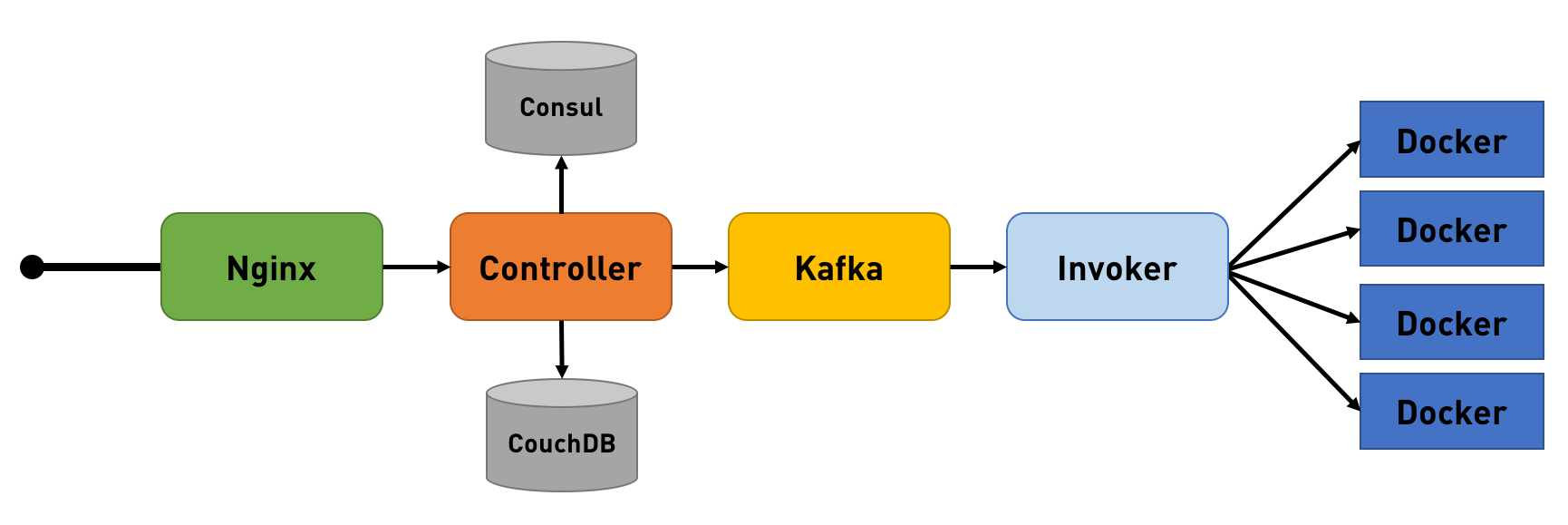}
\caption{Openwhisk architecture\cite{behindopenwhisk}}\label{fig:openwhisk}
\end{figure}

IBM has open-sourced its serverless compute service to the Apache Software
Foundation and is among the first to publicly conjoin the serverless paradigm
with a cloud-native infrastructure for distributed mobile application
architectures in \cite{cloudnativeeventbased}. Figure~\ref{fig:openwhisk}
outlines the components of the platform:

\begin{itemize}
  \item Nginx as a web request front-end reverse proxy that terminates user
  connections and translates RESTful requests into API calls of the controller
  \item Controller written in Scala that asserts entitlement and orchestrates
  function executions by sorting events into invoker's message queues
  \item Kafka to buffer and persist the events and ensure delivery
  \item Invokers that process events by loading
  function code from CouchDB, executing it in a Docker container and storing the
  result back to CouchDB
  \item Consul to manage the platform's container infrastructure (inventory
  management) of proxy, controller, database and worker containers
  \item CouchDB that maintains security policies, authentication data, function
  code, execution quotas, execution results, a.o.
\end{itemize}

The OpenWhisk architecture design uses reverse proxy load balancing (Nginx),
master-worker job scheduling (controller-invoker) with persistent, distributed
queuing (Kafka). These are typical best practices for scalable web application
architectures. Notably, the application state and platform state are held
separately in Consul and CouchDB, a key-value store and a document-based store,
both distributed in nature but with different replication protocols.
Event dispatching is performed in two stages by the reverse proxy and the
controller that selects a worker's queue. The default controller implementation
only uses round-robin but the architecture is suitable to implement more
sophisticated event scheduling algorithms. Kafka uses distributed and replicated
event message queues for reliability. Its coordination framework employs
consensus protocols to determine ownership (responsibility) and replication
(reliability) of queue partitions, which allows the controller to be implemented
as a stateless component. However, there is currently no mechanism to consider
data locality or context caching when dispatching an event.

\subsubsection{OpenFaaS}

More recently, Alex Ellis came out a winner of a Docker contest (Moby's cool
hacks) with a serverless architecture that reuses the Cloud Native Computing
Foundation's components Kubernetes (container orchestration), Prometheus
(monitoring database) and Docker (container runtime) - usually known for
microservice architectures - in a Function as a Service platform
architecture\cite{openfaas}:

\begin{itemize}
  \item Gateway written in Go to dispatch events to running containers 
  \item Worker containers that can run a certain function type
  \item Registry of container images
\end{itemize}

The	gateway monitors the container cluster resource metrics (through
Prometheus) to scale the number of worker containers. Functions are provided as
container images and at least one instance is required each to run for
event dispatching. OpenFaaS maps the serverless model (functions) one-to-one
onto containers and uses the container platform to track state of resources and
services. Multiple invocations of a function run concurrently in a container
and are isolated by different processes only.

\subsubsection{Serverless Prototype}

Garrett McGrath from University of Notre Dame, Indiana \cite{McGrath} designed a
prototype based on Microsoft Azure services. It uses load-balanced RESTful
WebServices to accept and classify a function request to create a task called
"execution request object" that contains the function meta-data (context) and
inputs (event). Instead of submitting the job through the messaging layer,
workers promote their availability through the message layer. A worker may
either have a container running that has the function code loaded (warm) or
unallocated memory to launch a new container (cold). Idle function containers
promote their availability in per-function queues (warm queues), workers promote
their capacity separately (cold queue). As explained in \cite{McGrath}, an event
is dispatched to the first in a queue of available containers per function
regardless the available context data.
\section{Objectives}\label{sec:objectives}

Serverless allows to dispatch the invocation of individual functions to any
resource node. This provides a new degree of flexibility alongside data
replication and placement which allows the platform to scale the execution of an
application comprised of functions across resources that also hold the
related data but every invocation may come with some overhead.

The main objective of this thesis is to design and evaluate an adaptive event
dispatching to increase serverless function execution resource efficiency.

To assess the potential for improvement, the design of event dispatching in
existing open source serverless frameworks has to be reviewed. Initial survey of
the matter has asserted that open serverless system designs do not optimize for
function execution time and it has revealed that public serverless providers
might be primarily concerned with resource utilization rather than computing
efficiency.

Under the hypothesis that execution overhead is dominated by both dispatching
latency and fetching execution context, a plan shall be designed to generate
application knowledge on event-, function- and data dependecies and to
use this knowledge to optimize platform resource efficiency. The target is to
adaptively improve event dispatching for resource efficiency at runtime.

The objectives comprise the analysis of (at least) one serverless framework, a
thorough evaluation of design options for serverless event dispatching, the
implementation of an adaptive event dispatcher and its evaluation in a
serverless testbed.

\section{Methods}\label{sec:methods}

In the following, methods for the design of adaptive event dispatching are
presented that may be applied to serverless. Advancements in other fields such
as Big Data analytics and many-task computing (MTC) and the general area of distributed
computing (Grid computing) have been considered.

\subsection{Data-awareness}\label{sec:dataawareness}
In the context of Grid computing, Ranganathan and
Foster\cite{decouplingcomputedata} have shown for synthetic, data-intensive
workloads that \enquote{scheduling jobs to idle processors, and then moving data
if required, performs significantly less well than algorithms that also consider
data location when scheduling} and they \enquote{achieve particularly good
performance with an approach in which jobs are always scheduled where data is
located, and a separate replication process at each site periodically generates
new replicas of popular datasets}. Their scheduling architecture is a two-stage
hierarchical system (according to the Grid infrastructure) that (1) forwards
jobs to sites that contain the data and (2) replicate popular data to meet
computing demand with local data access. These two aspects of data-awareness are
also popular with data-intensive computing (Big Data), which partitions,
randomly distributes, and replicates ingested data in order to maximize resource
utilization during parallel processing. However, disconnected strategies may
unfortunately break this overall system property. As discussed in
CoLoc\cite{coloc}, decoupling the Hadoop data store (HDFS) and container-based
worker pools (YARN) requires a data-aware container placement to keep data
accesses local, and can reduce execution time up to 35\% as compared to default
container scheduling and HDFS block placement.
Serverless application design faces the same issue when decoupling the program
state from the execution. The necessary design of a data-aware serverless
event dispatching can either reactively or proactively localize data:
\begin{itemize}
  \item \textbf{reactive localization}: event dispatching looks up data
  locations and schedules processing on optimal locations
  \item \textbf{proactive localization}: event dispatching clusters events and
  proactively choses locations considering that changes may cause
  replication or migration of data
\end{itemize}
Serverless computing however would be different from data-intensive computing in
that it also comprises tasks with fewer data access and more complex data
dependencies that vary with every event. Many-task computing as presented in
\cite{matrix} addresses a broader range of applications with execution time
granularities of 64ms to 8 seconds and both communication-intensive as well as
data-intensive tasks, which fits with serverless granularity. It needs to be
assessed whether these MTC advancements on data-awareness\cite{matrixsched} can
be applied to serverless adaptive event dispatching.

\subsection{Distributed scheduling}\label{sec:distributedscheduling}
Recent work in many-task computing, like MATRIX \cite{matrix}\cite{matrixsched}
or Sparrow \cite{sparrow} suggests that the trend for over-decomposition of Big
Data applications requires a decentralized or fully-distributed task scheduling
to achieve the required scheduling throughput. The serverless trend towards
functions faces a similar over-decomposition and latency has already been
identified an issue\cite{enddominance}\cite{nextgencloud}. Any centralized
approach can be clearly dismissed for scalability issues, so scheduling requires
some coordination over shared (externalized) system state.

Current serverless follows the cloud-native design\cite{cloudnative} to use
decentralized pools of service instances dedicated to load balancing (cmp.
controller \cite{ibm}, gateway \cite{openfaas}) to scale the front tier.
OpenWhisk\cite{ibm} uses Kafka as reliable distributed message queuing service
to persist tasks on message queueing servers. Opposingly, the fully-distributed
approach developed by Wang et al \cite{matrix}\cite{matrixsched} spreads queued
tasks across all system nodes which has the benefit to be able to dequeue and
process tasks locally. It uses a distributed hash table (ZHT\cite{zht}) to store
tasks, data dependency and data locality but employs randomized neighbor
selection for work stealing, that has been questioned in \cite{albatross} to
potentially cause poor utilization and scalability in some scenarios.
Instead, it suggests distributed message queuing over its distributed hash
table, i.e. across all nodes. It is sensible to adopt the fully-distributed
approach as locally spawned events could be quickly processed if the node also
hosts the required data or if the data access to computation ratio is low.

\subsection{Event clustering}\label{sec:eventclustering}

Serverless event executions can be data-independent or data-coupled, sequential
(flow chain) or parallel, recurrent or unique, small or large (w.r.t. memory),
single- or multithreaded, short or long. Furthermore, serverless event
dispatching needs to consider the system configuration, the data locality and
resource availability to achieve the highest resource efficiency. The problem is
how to use these parameters to infer the optimal location for an arriving event
and, in case of proactive localization, how to cluster similar events.

OpenWhisk\cite{behindopenwhisk} and OpenFaaS\cite{openfaas} use by default
round-robin and scale simply by machine load. Data-aware
MATRIX\cite{matrixsched} uses task graph dependencies and data dependencies to
schedule on optimal locations and employs work-stealing to balance load.
Considering the directed acyclic graph of a job's workflow is also common for
modern distributed computing platforms (cmp. Apache Spark). An event typically
carries the name of the function to be invoked, so it can be used to cluster
events by the code they trigger. An event may also carry function arguments. The
values might give hints to which data sets might be required to execute the
event. Eventually, metadata about an event can be used, e.g. the event source
that initiated the request. Depending on the code (function name), the referred
data (arguments) and the metadata (origin, endpoints, proxies) it might be
possible to cluster events and direct groups to an optimal worker that has most
of the required execution context in memory. For serverless, a common metric is
required to describe optimality of a location which exceeds the current
data-aware approaches\cite{matrixsched}\cite{coloc}.

\section{Project timeline}\label{sec:timeline}

\begin{figure}[H]
\centering
\includegraphics[width=\textwidth]{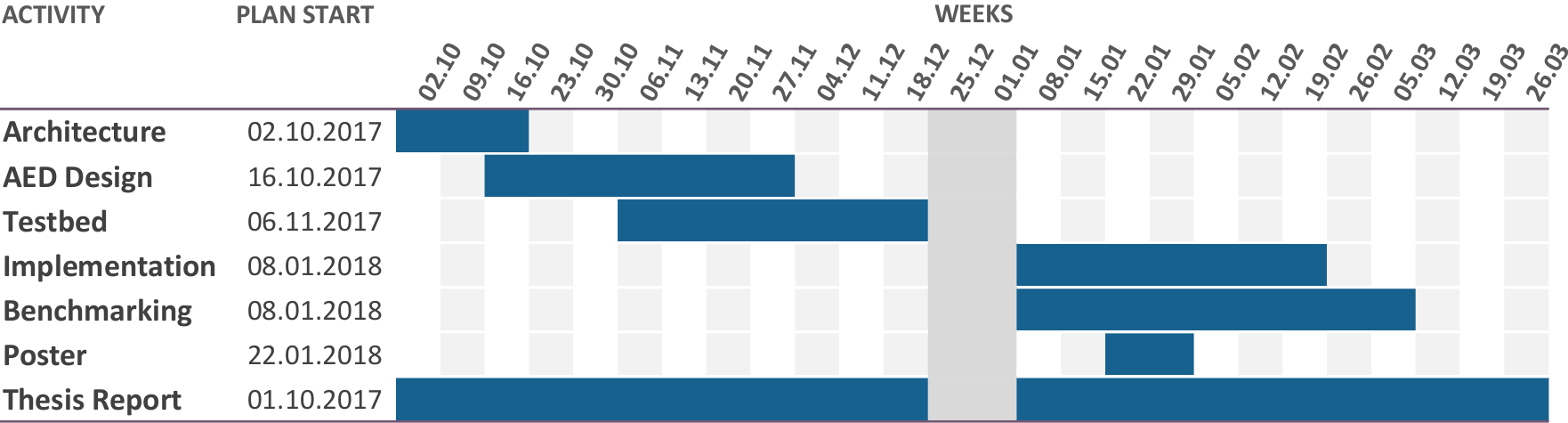}
\caption{Project timeline}\label{fig:timeline}
\end{figure}

The project timeline is set from October 2017 until the submission of the
dissertation in March 2018 and is diplayed in figure~\ref{fig:timeline}. The
task plan is result-oriented, i.e. tasks are structured to deliver intermediate
results. All results feed into the thesis reporting which spans the entire
project time and produces the final dissertation.

The project uses the waterfall model to engineer a solution. It starts with the
choice and description of the serverless platform (or simulator) to integrate
with. Architectures have been investigated during the initial survey, so it is
expected to be finished within the first three weeks.
The design of adaptive event dispatching follows, which provides a specification
of the envisioned adaptive event dispatching subsystem. In parallel, the setup
of a testbed should be promoted to assure that hardware is available and the
chosen serverless platform software can be setup before the end of year.
Beginning next year, the implementation and benchmarking efforts kick off to
evaluate the platform's default event dispatching and to implement the designed
mechanism until final evaluation of the improvements. A separate task end of
January 2018 is preserved to prepare a poster demonstration. The final three
weeks of the project are left for dissertation writing and to mitigate the risk
of accumulating delays.
\section{Deliverables}\label{sec:deliverables}

This project is laid out to design and implement an adaptive event dispatching
for serverless platforms. Hence, the system design, the implementation and an
evaluation should be deliverables of this project. For documentation of the
project's progress, a poster is to be presented beginning of February. The final
dissertation should be issued end of March.

The \textbf{design} comprises the choice of a serverless platform architecture
as well as the design of the adaptive event dispatching subsystem. The resulting
specification describes the components and mechanisms of adaptive event
dispatching and interworking with the platform.

The \textbf{implementation} is a proof-of-concept realization of the designed
specification and may use discrete event simulation in case that hardware
resources are not available at the time to integrate the developed mechanism
with a serverless platform.

This work thesisizes that intelligent dispatching can improve cache hits on
colocated data caches/stores to reduce function execution time and resource
consumption. The \textbf{evaluation} should assess the quality and efficiency of
the designed system. If possible, the evaluation should also compare against one 
existing serverless plaform. The quality of event dispatching can be assessed
comparing the actual execution time of a task and its ideal execution time. The
efficiency of the system is the proportion of time that a system spends actually
executing tasks, while the remaining time is used to move data, wait for
communication, etc. System utilization may complement the evaluation to show
system state. (cmp. \cite{matrixsched})

\setstretch{1}
\bibliographystyle{plainnat}
\bibliography{interim-report}

\begin{thebibliography}{46}
\providecommand{\natexlab}[1]{#1}
\providecommand{\url}[1]{\texttt{#1}}
\expandafter\ifx\csname urlstyle\endcsname\relax
  \providecommand{\doi}[1]{doi: #1}\else
  \providecommand{\doi}{doi: \begingroup \urlstyle{rm}\Url}\fi

\bibitem[cnc(2015)]{cncf}
Charter - {Cloud} {Native} {Computing} {Foundation}, November 2015.
\newblock URL \url{https://www.cncf.io/about/charter/}.

\bibitem[ama(2017{\natexlab{a}})]{amazon}
{AWS Lambda: How It Works}, September 2017{\natexlab{a}}.
\newblock URL
  \url{http://docs.aws.amazon.com/lambda/latest/dg/lambda-introduction.html}.

\bibitem[ama(2017{\natexlab{b}})]{amazoniot}
{How the AWS IoT Platform Works}, September 2017{\natexlab{b}}.
\newblock URL \url{https://aws.amazon.com/iot-platform/how-it-works/}.

\bibitem[ama(2017{\natexlab{c}})]{amazonpricing}
{AWS Lambda Pricing}, September 2017{\natexlab{c}}.
\newblock URL \url{https://aws.amazon.com/lambda/pricing/}.

\bibitem[azu(2017{\natexlab{a}})]{azure}
{Azure Functions Overview}, September 2017{\natexlab{a}}.
\newblock URL
  \url{https://docs.microsoft.com/en-us/azure/azure-functions/functions-overview}.

\bibitem[azu(2017{\natexlab{b}})]{azurepricing}
{Azure Functions pricing}, September 2017{\natexlab{b}}.
\newblock URL
  \url{https://azure.microsoft.com/en-us/pricing/details/functions/}.

\bibitem[dat(2017)]{databricks}
{Databricks Serverless: Next Generation Resource Management for Apache Spark},
  September 2017.
\newblock URL
  \url{https://databricks.com/blog/2017/06/07/databricks-serverless-next-generation-resource-management-for-apache-spark.html}.

\bibitem[goo(2017{\natexlab{a}})]{google}
{Cloud Functions Overview}, September 2017{\natexlab{a}}.
\newblock URL \url{https://cloud.google.com/functions/docs/concepts/overview}.

\bibitem[goo(2017{\natexlab{b}})]{googlepricing}
{Google Cloud Functions Pricing}, September 2017{\natexlab{b}}.
\newblock URL \url{https://cloud.google.com/functions/pricing}.

\bibitem[ibm(2017{\natexlab{a}})]{ibm}
{About Cloud Functions}, September 2017{\natexlab{a}}.
\newblock URL
  \url{https://console.bluemix.net/docs/openwhisk/openwhisk_about.html#about-cloud-functions}.

\bibitem[ibm(2017{\natexlab{b}})]{ibmpricing}
{IBM Cloud Functions pricing}, September 2017{\natexlab{b}}.
\newblock URL \url{https://console.bluemix.net/openwhisk/learn/pricing}.

\bibitem[ope(2017{\natexlab{a}})]{openlambdalb}
Openlambda dev meeting july 5, Jul 2017{\natexlab{a}}.
\newblock URL \url{http://open-lambda.org/resources/slides/july-5-16.pdf}.

\bibitem[ope(2017{\natexlab{b}})]{openlambdaslides}
Serverless computation with openlambda, June 2017{\natexlab{b}}.
\newblock URL
  \url{http://open-lambda.org/resources/slides/ol-first-meeting.pdf}.

\bibitem[Adzic and Chatley(2017)]{economicimpact}
Gojko Adzic and Robert Chatley.
\newblock Serverless computing: Economic and architectural impact.
\newblock In \emph{Proceedings of the 2017 11th Joint Meeting on Foundations of
  Software Engineering}, ESEC/FSE 2017, pages 884--889, New York, NY, USA,
  2017. ACM.
\newblock ISBN 978-1-4503-5105-8.
\newblock \doi{10.1145/3106237.3117767}.
\newblock URL \url{http://doi.acm.org/10.1145/3106237.3117767}.

\bibitem[Bainomugisha et~al.(2013)Bainomugisha, Carreton, Cutsem, Mostinckx,
  and Meuter]{reactiveprogrammingsurvey}
Engineer Bainomugisha, Andoni~Lombide Carreton, Tom~van Cutsem, Stijn
  Mostinckx, and Wolfgang~de Meuter.
\newblock A survey on reactive programming.
\newblock \emph{ACM Comput. Surv.}, 45\penalty0 (4):\penalty0 52:1--52:34,
  August 2013.
\newblock ISSN 0360-0300.
\newblock \doi{10.1145/2501654.2501666}.
\newblock URL \url{http://doi.acm.org/10.1145/2501654.2501666}.

\bibitem[Baldini et~al.(2016)Baldini, Castro, Cheng, Fink, Ishakian, Mitchell,
  Muthusamy, Rabbah, and Suter]{cloudnativeeventbased}
Ioana Baldini, Paul Castro, Perry Cheng, Stephen Fink, Vatche Ishakian, Nick
  Mitchell, Vinod Muthusamy, Rodric Rabbah, and Philippe Suter.
\newblock Cloud-native, event-based programming for mobile applications.
\newblock In \emph{Proceedings of the International Conference on Mobile
  Software Engineering and Systems}, MOBILESoft '16, pages 287--288, New York,
  NY, USA, 2016. ACM.
\newblock ISBN 978-1-4503-4178-3.
\newblock \doi{10.1145/2897073.2897713}.
\newblock URL \url{http://doi.acm.org/10.1145/2897073.2897713}.

\bibitem[Barcelo et~al.(2015)Barcelo, Llorca, Tulino, and Raman]{jaime}
M.~Barcelo, J.~Llorca, A.~M. Tulino, and N.~Raman.
\newblock The cloud service distribution problem in distributed cloud networks.
\newblock In \emph{2015 IEEE International Conference on Communications (ICC)},
  pages 344--350, June 2015.
\newblock \doi{10.1109/ICC.2015.7248345}.

\bibitem[Barr(2015)]{carbonamazon}
Jeff Barr, June 2015.
\newblock URL
  \url{https://aws.amazon.com/blogs/aws/cloud-computing-server-utilization-the-environment/}.

\bibitem[Boyd(2016)]{serverlessnoops}
Mark Boyd.
\newblock {The Road to NoOps: Serverless Computing is Quickly Gaining
  Momentum}, May 2016.
\newblock URL
  \url{https://thenewstack.io/serverless-computing-growing-quickly/}.

\bibitem[Boyd(2017)]{iotlearn}
Mark Boyd.
\newblock What serverless and the internet of things can learn from each other,
  May 2017.
\newblock URL \url{https://thenewstack.io/iot-serverless-can-learn/}.

\bibitem[Cohen et~al.(2015)Cohen, Lewin-Eytan, Naor, and Raz]{danny}
R.~Cohen, L.~Lewin-Eytan, J.~S. Naor, and D.~Raz.
\newblock Near optimal placement of virtual network functions.
\newblock In \emph{2015 IEEE Conference on Computer Communications (INFOCOM)},
  pages 1346--1354, Apr 2015.
\newblock \doi{10.1109/INFOCOM.2015.7218511}.

\bibitem[Eivy(2017)]{Economics}
A.~Eivy.
\newblock Be wary of the economics of serverless cloud computing.
\newblock \emph{IEEE Cloud Computing}, 4\penalty0 (2):\penalty0 6--12, Mar
  2017.
\newblock ISSN 2325-6095.
\newblock \doi{10.1109/MCC.2017.32}.

\bibitem[Ellis(2017)]{openfaas}
Alex Ellis, Jan 2017.
\newblock URL \url{https://blog.alexellis.io/functions-as-a-service/}.

\bibitem[Fox et~al.(2017)Fox, Ishakian, Muthusamy, and Slominski]{wosc}
Geoffrey~Charles Fox, Vatche Ishakian, Vinod Muthusamy, and Aleksander
  Slominski.
\newblock Status of serverless computing and {Function-as-a-Service} ({FaaS})
  in industry and research.
\newblock Technical report, First International Workshop on Serverless
  Computing (WoSC) 2017, 2017.

\bibitem[Glikson et~al.(2017)Glikson, Nastic, and Dustdar]{devicelessedge}
Alex Glikson, Stefan Nastic, and Schahram Dustdar.
\newblock Deviceless edge computing: Extending serverless computing to the edge
  of the network.
\newblock In \emph{Proceedings of the 10th ACM International Systems and
  Storage Conference}, SYSTOR '17, pages 28:1--28:1, New York, NY, USA, 2017.
  ACM.
\newblock ISBN 978-1-4503-5035-8.
\newblock \doi{10.1145/3078468.3078497}.
\newblock URL \url{http://doi.acm.org/10.1145/3078468.3078497}.

\bibitem[Gualtieri(2011)]{noops}
Mike Gualtieri, February 2011.
\newblock URL
  \url{https://go.forrester.com/blogs/11-02-07-i_dont_want_devops_i_want_noops/}.

\bibitem[Hendrickson et~al.(2016)Hendrickson, Sturdevant, Harter,
  Venkataramani, Arpaci-Dusseau, and Arpaci-Dusseau]{openlambda}
Scott Hendrickson, Stephen Sturdevant, Tyler Harter, Venkateshwaran
  Venkataramani, Andrea~C. Arpaci-Dusseau, and Remzi~H. Arpaci-Dusseau.
\newblock Serverless computation with openlambda.
\newblock In \emph{8th {USENIX} Workshop on Hot Topics in Cloud Computing
  (HotCloud 16)}, Denver, CO, 2016. {USENIX} Association.
\newblock URL
  \url{https://www.usenix.org/conference/hotcloud16/workshop-program/presentation/hendrickson}.

\bibitem[Kepes(2015)]{comatose}
Ben Kepes, June 2015.
\newblock URL
  \url{https://www.forbes.com/sites/benkepes/2015/06/03/30-of-servers-are-sitting-comatose-according-to-research/#223dbc6c59c7}.

\bibitem[Koller and Williams(2017)]{enddominance}
Ricardo Koller and Dan Williams.
\newblock Will serverless end the dominance of linux in the cloud?
\newblock In \emph{Proceedings of the 16th Workshop on Hot Topics in Operating
  Systems}, HotOS '17, pages 169--173, New York, NY, USA, 2017. ACM.
\newblock ISBN 978-1-4503-5068-6.
\newblock \doi{10.1145/3102980.3103008}.
\newblock URL \url{http://doi.acm.org/10.1145/3102980.3103008}.

\bibitem[Lee and Lee(1985)]{harmonic}
C.~C. Lee and D.~T. Lee.
\newblock A simple on-line bin-packing algorithm.
\newblock \emph{J. ACM}, 32\penalty0 (3):\penalty0 562--572, Jul 1985.
\newblock ISSN 0004-5411.
\newblock \doi{10.1145/3828.3833}.
\newblock URL \url{http://doi.acm.org/10.1145/3828.3833}.

\bibitem[Li et~al.(2013)Li, Zhou, Brandstatter, Zhao, Wang, Rajendran, Zhang,
  and Raicu]{zht}
T.~Li, X.~Zhou, K.~Brandstatter, D.~Zhao, K.~Wang, A.~Rajendran, Z.~Zhang, and
  I.~Raicu.
\newblock {ZHT}: A light-weight reliable persistent dynamic scalable zero-hop
  distributed hash table.
\newblock In \emph{2013 IEEE 27th International Symposium on Parallel and
  Distributed Processing}, pages 775--787, May 2013.
\newblock \doi{10.1109/IPDPS.2013.110}.

\bibitem[Mann(2015)]{placementsurvey}
Zolt\'{a}n~\'{A}d\'{a}m Mann.
\newblock Allocation of virtual machines in cloud data centers - a survey of
  problem models and optimization algorithms.
\newblock \emph{ACM Comput. Surv.}, 48\penalty0 (1):\penalty0 11:1--11:34,
  August 2015.
\newblock ISSN 0360-0300.
\newblock \doi{10.1145/2797211}.
\newblock URL \url{http://doi.acm.org/10.1145/2797211}.

\bibitem[McGrath and Brenner(2017)]{McGrath}
G.~McGrath and P.~R. Brenner.
\newblock Serverless computing: Design, implementation, and performance.
\newblock In \emph{2017 IEEE 37th International Conference on Distributed
  Computing Systems Workshops (ICDCSW)}, pages 405--410, June 2017.
\newblock \doi{10.1109/ICDCSW.2017.36}.

\bibitem[McGrath et~al.(2016)McGrath, Short, Ennis, Judson, and
  Brenner]{cloudeventprogramming}
G.~McGrath, J.~Short, S.~Ennis, B.~Judson, and P.~Brenner.
\newblock Cloud event programming paradigms: Applications and analysis.
\newblock In \emph{2016 IEEE 9th International Conference on Cloud Computing
  (CLOUD)}, pages 400--406, June 2016.
\newblock \doi{10.1109/CLOUD.2016.0060}.

\bibitem[Mell and Grance(2011)]{nist}
Peter~M. Mell and Timothy Grance.
\newblock {SP 800-145. The NIST Definition of Cloud Computing}.
\newblock Technical report, National Institute of Standards \& Technology,
  Gaithersburg, MD, United States, 2011.

\bibitem[MSV(2017)]{behindopenwhisk}
Janakiram MSV, Feb 2017.
\newblock URL
  \url{https://thenewstack.io/behind-scenes-apache-openwhisk-serverless-platform/}.

\bibitem[orrwaws(2016)]{singlethreaded}
orrwaws, Sep 2016.
\newblock URL \url{https://forums.aws.amazon.com/thread.jspa?messageID=708378}.

\bibitem[Ousterhout et~al.(2013)Ousterhout, Wendell, Zaharia, and
  Stoica]{sparrow}
Kay Ousterhout, Patrick Wendell, Matei Zaharia, and Ion Stoica.
\newblock Sparrow: Distributed, low latency scheduling.
\newblock In \emph{Proceedings of the Twenty-Fourth ACM Symposium on Operating
  Systems Principles}, SOSP '13, pages 69--84, New York, NY, USA, 2013. ACM.
\newblock ISBN 978-1-4503-2388-8.
\newblock \doi{10.1145/2517349.2522716}.
\newblock URL \url{http://doi.acm.org/10.1145/2517349.2522716}.

\bibitem[Ranganathan and Foster(2002)]{decouplingcomputedata}
K.~Ranganathan and I.~Foster.
\newblock Decoupling computation and data scheduling in distributed
  data-intensive applications.
\newblock In \emph{Proceedings 11th IEEE International Symposium on High
  Performance Distributed Computing}, pages 352--358, 2002.
\newblock \doi{10.1109/HPDC.2002.1029935}.

\bibitem[Renner et~al.(2016)Renner, Thamsen, and Kao]{coloc}
T.~Renner, L.~Thamsen, and O.~Kao.
\newblock Coloc: Distributed data and container colocation for data-intensive
  applications.
\newblock In \emph{2016 IEEE International Conference on Big Data (Big Data)},
  pages 3008--3015, Dec 2016.
\newblock \doi{10.1109/BigData.2016.7840954}.

\bibitem[Sadooghi et~al.(2016)Sadooghi, Kumar, Wang, Zhao, Li, and
  Raicu]{albatross}
I.~Sadooghi, G.~Kumar, K.~Wang, D.~Zhao, T.~Li, and I.~Raicu.
\newblock Albatross: An efficient cloud-enabled task scheduling and execution
  framework using distributed message queues.
\newblock In \emph{2016 IEEE 12th International Conference on e-Science
  (e-Science)}, pages 11--20, Oct 2016.
\newblock \doi{10.1109/eScience.2016.7870881}.

\bibitem[Schmidt et~al.(2008)Schmidt, Anicic, and
  St{\"u}hmer]{eventdrivenreactivity}
Kay-Uwe Schmidt, Darko Anicic, and Roland St{\"u}hmer.
\newblock Event-driven reactivity: A survey and requirements analysis.
\newblock In \emph{SBPM2008: 3rd international Workshop on Semantic Business
  Process Management in conjunction with the 5th European Semantic Web
  Conference (ESWC'08)}. CEUR Workshop Proceedings (CEUR-WS.org, ISSN
  1613-0073), June 2008.
\newblock URL \url{http://sbpm2008.fzi.de/paper/paper7.pdf}.

\bibitem[Toffetti et~al.(2017)Toffetti, Brunner, Bl{\"o}chlinger, Spillner, and
  Bohnert]{cloudnative}
Giovanni Toffetti, Sandro Brunner, Martin Bl{\"o}chlinger, Josef Spillner, and
  Thomas~Michael Bohnert.
\newblock Self-managing cloud-native applications: Design, implementation, and
  experience.
\newblock \emph{Future Generation Computer Systems}, 72\penalty0 (Supplement
  C):\penalty0 165 -- 179, 2017.
\newblock ISSN 0167-739X.
\newblock \doi{https://doi.org/10.1016/j.future.2016.09.002}.
\newblock URL
  \url{http://www.sciencedirect.com/science/article/pii/S0167739X16302977}.

\bibitem[Varghese and Buyya(2017)]{nextgencloud}
Blesson Varghese and Rajkumar Buyya.
\newblock Next generation cloud computing: New trends and research directions.
\newblock \emph{CoRR}, abs/1707.07452, 2017.
\newblock URL \url{http://arxiv.org/abs/1707.07452}.

\bibitem[Wang et~al.(2014)Wang, Zhou, Li, Zhao, Lang, and Raicu]{matrixsched}
K.~Wang, X.~Zhou, T.~Li, D.~Zhao, M.~Lang, and I.~Raicu.
\newblock Optimizing load balancing and data-locality with data-aware
  scheduling.
\newblock In \emph{2014 IEEE International Conference on Big Data (Big Data)},
  pages 119--128, Oct 2014.
\newblock \doi{10.1109/BigData.2014.7004220}.

\bibitem[Wang et~al.(2013)Wang, Brandstatter, and Raicu]{matrix}
Ke~Wang, Kevin Brandstatter, and Ioan Raicu.
\newblock Simmatrix: Simulator for many-task computing execution fabric at
  exascale.
\newblock In \emph{Proceedings of the High Performance Computing Symposium},
  HPC '13, pages 9:1--9:9, San Diego, CA, USA, 2013. Society for Computer
  Simulation International.
\newblock ISBN 978-1-62748-033-8.
\newblock URL \url{http://dl.acm.org/citation.cfm?id=2499968.2499977}.

\end{thebibliography}

\end{document}